\pdfoutput=1
\documentclass[10pt,twocolumn,letterpaper]{article}

\usepackage{cvpr}
\usepackage{times}
\usepackage{epsfig}
\usepackage{graphicx}
\usepackage{amsmath}
\usepackage{amssymb}
\usepackage{appendix}
\usepackage{listings}

\cvprfinalcopy

\begin{document}  

\title{AudioSetMix: Enhancing Audio-Language Datasets with LLM-Assisted Augmentations}

\author{David Xu\\
Princeton University\\
{\tt\small dx8527@princeton.edu}
}

\maketitle

\begin{abstract}
Multi-modal learning in the audio-language domain has seen significant advancements in recent years. However, audio-language learning faces challenges due to limited and lower-quality data compared to image-language tasks. Existing audio-language datasets are notably smaller, and manual labeling is hindered by the need to listen to entire audio clips for accurate labeling.

Our method systematically generates audio-caption pairs by augmenting audio clips with natural language labels and corresponding audio signal processing operations. Leveraging a Large Language Model, we generate descriptions of augmented audio clips with a prompt template. This scalable method produces AudioSetMix, a high-quality training dataset for text-and-audio related models.

Integration of our dataset improves models performance on benchmarks by providing diversified and better-aligned examples. Notably, our dataset addresses the absence of modifiers (adjectives and adverbs) in existing datasets. By enabling models to learn these concepts, and generating hard negative examples during training, we achieve state-of-the-art performance on multiple benchmarks.

\end{abstract}

\section{Introduction}

In recent years, there has been a large amount of work in expanding comprehension of audio content by augmenting audio signal with information from another modality such as natural language. Tasks such as text-to-audio generation (TTA) ~\cite{audioldm} ~\cite{audioldm2}, text-guided audio editing ~\cite{audit}, automatic audio-captioning ~\cite{Mei_captioning} ~\cite{Gontier}, and text-to-audio retrieval ~\cite{mei2022metric}~\cite{Koepke_2023} have been proposed as objectives that improve model understanding of audio signal ~\cite{wavcaps}.

A closely related field to audio-language learning is the vision-language learning, which comprises of tasks such as visual question-answering ~\cite{vqa} and text-to-image generation ~\cite{t2i} ~\cite{stable_diffusion}.  However, unlike audio-language, the vision-language learning benefits from the existence of large-scale, high quality datasets such as MSCOCO ~\cite{mscoco}. This makes it possible to pretrain large, powerful models to learn vision-language multimodal embeddings, which can then be applied for downstream tasks ~\cite{oscar} ~\cite{blip} ~\cite{stable_diffusion}. On the other hand, a significant challenge of audio-language learning is the lack of a large dataset consisting of high-quality audio-caption pairs. We note the distinction between captioned natural sound and captioned speech, the latter of which is more readily available. A commonly used dataset in audio-language is the AudioSet dataset ~\cite{audioset}, a collection of 2M YouTube videos of natural sounds with multi-label annotations. However, it is quite clear that labels alone are not sufficient to replace high-quality audio captions. Furthermore, datasets such as AudioCaps ~\cite{audiocaps} that provide human-generated text captions for audio are generally not large enough to train a deep neural network, and only suitable for fine-tuning ~\cite{wavcaps}.

In this work, we introduce AudioSetMix, an audio-caption dataset generated through the application of audio transformations to clips from AudioSet. In addition, we use prompt engineering and large language model (LLM) to ensure that the transformed audio and its caption are aligned. Our dataset supports speed, pitch, volume, and duration augmentations for individual clips, as well as mixing and concatenation augmentations to combine multiple clips into one. Besides having high quality text descriptions for supervised audio-language tasks, our data augmentation scheme also supports a dataset for studying text-guided audio manipulations, as we have access to both the original and edited audio.

To demonstrate the effectiveness of our dataset, we train a state-of-the-art model from the 2022/2023 DCASE Challenge on Language-Based Audio Retrieval (Task 6B) ~\cite{dcase} using AudioSetMix. We demonstrate that our model exhibits an improved understanding of common audio event modifiers such as volume or duration, as well as a better retrieval score overall compared to baseline models. Finally, we introduce a hard negative mining technique for the AudioSetMix data which further boosts model performance.

\section{Background and Related Work}

\subsection{Dataset Improvements for Audio-Language Learning}
In light of the dataset shortcomings for audio-language tasks, several workarounds for the data shortage have been proposed to train large models for audio-language tasks such as TTA. These approaches can be broadly classified into three categories. 

The first approach is to use predefined text templates to form rough approximations of descriptions. The simplest text templates are proposed by~\cite{audiogen}, in which audio labels are concatenated together in a random order.~\cite{audiogen} also proposes a data augmentation for the AudioSet dataset by mixing multiple audio samples together. The corresponding text caption is simply the concatenated labels of each sample. To allow for more complex relationships to be expressed in the captions,~\cite{diffsound} randomly inserts $\langle MASK \rangle$ tokens between labels in the hope that the model will learn to substitute in relational words.~\cite{makeanaudio} improves this approach by applying a set of common audio augmentations to AudioSet data, and associates each augmentation with a caption template. 

The second approach trains models in a self-supervised manner by using a pretrained CLAP model to embed audio and text to a shared latent space~\cite{audioldm}~\cite{musiclm}. During training, when no captions are available, the CLAP model is used to perform a form of zero-shot audio captioning by substituting the text embedding with the audio embedding. This approach has been applied to both TTA and to music generation with good results. In particular, this approach has been adopted by AudioLDM~\cite{audioldm}, which achieves state-of-the-art performance on both TTA task and other audio editing tasks such as style transfer and audio inpainting.  

The third and most recent approach to overcome the description scarcity issue utilizes recent LLM models such as ChatGPT~\cite{gpt3} from OpenAI to generate text descriptions. This approach provides a few benefits. Firstly, text descriptions from LLM are much more varied when compared to text templates in terms of sentence structure and word choice. Our observations also indicate that LLM descriptions are also fairly realistic when compared to human descriptions. Secondly, the quality of captions from LLM can be readily improved using prompt engineering techniques such as few-shot or chain-of-thought prompting. Furthermore, using LLM allows us to increase the complexity of data augmentations without requiring an intractably large number of human created templates. 

The first work to adopt the third approach is WavCaps~\cite{wavcaps}, a large audio-captioned dataset using ChatGPT. WavCaps combines several weakly-labelled datasets by prompting ChatGPT to create a natural language caption, given a list of sound event labels. However, WavCaps does not incorporate any form of data augmentation in its captions, thus reducing the diversity and complexity of captions. The lack of data augmentation also hinders the training of text-guided audio editing models, as editing keywords do not appear often in AudioSet labels. Finally, WavCaps does not explore the idea of generating hard negative examples, such as including two audio-caption pairs that differ by only audio event modifiers. In the next section, we introduce AudioSetMix, an audio-caption dataset which addresses these concerns.
\begin{figure*}
    \centering
    \includegraphics[width=0.9\linewidth]{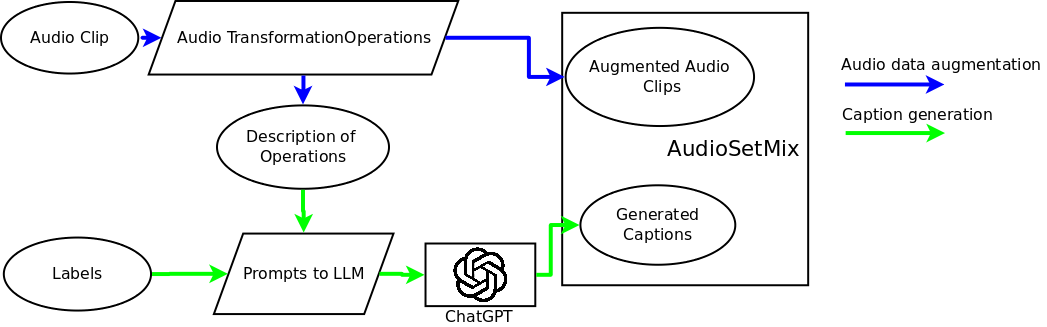}
    \caption{The overall pipeline to generate the AudioSetMix.}
    \label{fig:pipeline}
\end{figure*}

\section{AudioSetMix Dataset}\label{sec:datagen}
In this section, we describe the process of creating the AudioSetMix dataset. Firstly, we introduce the data source and its characteristics. Secondly we describe a four-stage pipeline for generating weak captions, including data preprocessing, audio clip augmentation, LLM-based caption generation, and postprocessing. Finally, we provide an analysis of AudioSetMix and compare it with existing audio-language datasets.

\subsection{TS-AudioSet }
The original AudioSet dataset consists of approximately 2 millions 10~second audio clips with human annotations. However, AudioSet labels are considered weak as an audio clip in its entirety may not correspond to its label due to interference such as background noise ~\cite{ts_audioset}. This imperfection is undesirable when training text-to-audio models as they may learn to associate labels with silence or white noise that dominate audio clips. For this reason, previous works ~\cite{regnet} restrict themselves to particular sound classes such as drumming. To resolve this issue, ~\cite{ts_audioset} released Temporally-Strong AudioSet~(TS-AudioSet), in which each audio clip has precise start and end timestamps and are labeled by humans. We use the clips and labels from TS-AudioSet as the basis for forming AudioSetMix.
\subsection{Data Generation}
We propose the following pipeline for generating clip captions using a LLM. We will first discuss our audio clip augmentations, then move into the details of generating audio-aligned captions that are specific to each augmentation method applied to the audio clips. Figure~\ref{fig:pipeline} illustrates the full data processing pipeline.

\subsubsection{Audio Data Preprocessing}
We first apply preprocessing to remove noisy or undesirable data. This largely consists of  filtering operations to audio clips and captions, referred to as duration-based filtering and class-based filtering respectively. Due to the noisiness of the raw audio-caption pairs, these filtering operations are needed to ensure the cleanliness of the data as the starting point for the subsequent enhancement and augmentation. For duration-based filtering, we remove clips with a duration of less than two seconds. This is because shorter clips tend to lack meaningful content and require long amounts of padding during training ~\cite{wavcaps}. Furthermore, short clips may be even further reduced if duration augmentations are applied in later stages. For class-based filtering, we remove clips with labels such as ``background/environment'' and ``unknown'' that lack semantically significant contents.
\begin{table*}[]
\centering
\caption{Math symbols used in describing data augmentation}
\begin{tabular}{p{0.15\linewidth} | p{0.8\linewidth} }
Math symbol & Description \\
 \hline
\hline
$A$ & A set of audio clips sampled from TS-AudioSet \\
$n$ & the number of clips in $A$, $n\in[1,5]$ \\
$T$ & A group of operations to transform the audio clips in $A$. Operations include Volume, Pitch, Speed, and Duration \\
$p_t$ & $p_t=0.3$ is the parameter of Bernoulli distribution that decides if a transformation $T_j\in T$ is applied to an audio clip  \\
$C_i$ & $C_i$ is an augmented audio clip, and $i \in [0, n]$ \\
$p_c$ & $p_c=0.2$ is the parameter of Bernoulli distribution that decides ``mixing'' operations for each consecutive pair of augmented clips $(C_{i-1}, C_i)$. Mixing operation can be either ``Concatenation'', or ``Mix'' \\
\end{tabular}
\label{tab:mathsym}
\end{table*}
\subsubsection{Audio Data Augmentation}
Before discussing in details about the audio data enrichment, Table~\ref{tab:mathsym} lists the math symbols used in this sub section to guide the reading. 

To generate a new audio-caption pair, we first select $n$  audio clips $A$ uniformly at random from TS-AudioSet. $n$ is also selected uniformly from range $[1, 5]$, as large $n$ cause the generated audio-caption pair to be unrealistic and overly complex.

Furthermore, we define a set of audio clip transformations $T$. The four types of transformations are volume, pitch, speed and duration, with implementation details of each transformation given below. Each transformation is also associated with a set of keywords which describe the transformation. For instance, the ``volume'' transformation is associated with keyword ``loud''.

\textbf{Volume}: Given clip C, we randomly apply either amplification or attenuation with uniformly random magnitude in range $[0.5, 1]$~dB to the entire clip. 

\textbf{Pitch}: Given clip C, we randomly shift the pitch by a uniformly random number of octaves between $[-0.5, 0.5]$. 

\textbf{Speed}: Given clip C, we randomly stretch C in time domain by uniformly random rate in $[0.8, 1.2]$. For this transformation, we use the \textit{TimeStretch()} function from \textit{torchaudio}, which preserves the pitch rate when modifying speed. 

\textbf{Duration}: Given clip C, we randomly reduce the length of C by half.

For each pair of audio clip $A_i$ and transformation $T_j$, we use a Bernoulli random variable with parameter $p_t$ to determine if $T_j$ should be applied to $A_i$. 

In addition, we define 2 transformations to combine two clips together:

\textbf{Concatenation}: Given clips $C_1$ and $C_2$, we concatenate $C_1$ and $C_2$ with 0.5~second of silence separating the clips.

\textbf{Mix}: Given clips $C_1$ and $C_2$, we randomly select a temporal offset for combining $C_1$ and $C_2$. We then draw a signal-to-noise ratio (SNR) between $[-5,5]$ and mix the two clips together.

 Once individual augmentations are applied to each clip, we combine the clips to form the final waveform. Let $[C_0, C_1, ...C_n]$ be the augmented clips. For every $i$, we combine clips $(C_{i-1}, C_i)$ using the mix transformation with probability $p_c$, or the concatenation transformation with probability $1-p_c$. In addition, $C_{i-1}$ and $C_i$ are said to occur simultaneously if they are mixed together. Otherwise, $C_{i-1}$ will occur before $C_i$.  We set $p_t = 0.3$ and $p_c = 0.2$ in our pipeline implementation.  The final clip is then padded/truncated such that the length of the clip is 10 seconds.

\subsubsection{Caption Generation}

\begin{table*}[]
\centering
\caption{Example LLM Query and Response. Using the original text labels, a query is constructed based on the operations in audio clips transformations. The basic prompt in the listing~\ref{lst:prompt} and query are sent to LLM to generate the response, which is the captions for the augmented audio clip.}
\begin{tabular}{p{0.2\textwidth} | p{0.36\textwidth} | p{0.36\textwidth}}
Original Labels & Query & Response\\
\hline
\hline
 1. \textcolor{red}{car driving}, \textcolor{blue}{firecracker}, \textcolor{green}{tree falling}

& 1. [{sound: \textcolor{red}{car driving}, description: \textbf{[loud]}, order: \textbf{0}}, {sound: \textcolor{blue}{firecracker}, description: \textbf{[background, high-pitch}], order:  \textbf{0}}, {sound: \textcolor{green}{tree falling}, description: \textbf{[fast]}, order: \textbf{1}}]

& 1. A car drove by loudly as the distant sound of a firecracker echoed in the background; suddenly, a tree fell with a swift crash.\\

2. \textcolor{red}{train horn}, \textcolor{blue}{fire engine/ fire truck (siren)}, \textcolor{green}{tree falling}

& 2. [{sound: \textcolor{red}{train horn}, description: \textbf{[loud]}, order: \textbf{0}}, {sound: \textcolor{blue}{fire engine/fire truck (siren)}, description: \textbf{[background, high-pitch]}, order: \textbf{1}}, {sound: \textcolor{green}{tree falling}, description: \textbf{[fast]}, order: \textbf{2}}] 
& 2. The blaring horn of a train resonated loudly, followed by the high-pitched wail of a fire engine's siren in the background, just before a tree toppled with a rapid crash.
\end{tabular}
\label{tab:query}
\end{table*}

Instead of using simple concatenation or other techniques in previous works~\cite{audiogen}~\cite{diffsound} to assemble the captions corresponding to the audio clip, we use LLM to generate natural language description of the new audio clips based on the augmentations applied in the previous step. In this section, we describe how we construct the prompts in order to generate text captions. 

To query the LLM, we introduce a JSON-formatted dictionary to describe each clip. The dictionary for clip $C$ contains the original list of labels from TS-AudioSet for $C$, the keywords for the transformations applied to $C$, and an $order$ value. We assign the first clip to have an $order = 0$. Furthermore, $C_{i-1}$ and $C_i$ will have the same $order$ value if they occur simultaneously, i.e. occurred as the consequence of 'mixing', otherwise, their $order$s will incrementally differ by 1. The final query to LLM is a list of the JSON dictionaries, as well as a prompt instructing LLM to generate a short, realistic description based on the dictionary values. 

\noindent\begin{minipage}{.45\textwidth}
\begin{lstlisting}[caption=Prompt for LLM,frame=tb,label=lst:prompt,breaklines=true]
I will provide a list of 
scenarios. For each scenario,
I want you to provide a 
short, one sentence story.
Each scenario will be 
described as a JSON list. 
Pay particular attention to 
the order attribute, which 
describes the temporal ordering
of the story. Only return 
the stories. 
\end{lstlisting}
\end{minipage}\hfill

The prompt for the LLM is illustrated in the Listing~\ref{lst:prompt}, and the detailed construction of the full queries to LLM and sample responses are shown in Table~\ref{tab:query}.

We select GPT 3.5 as our LLM of choice for the implementation due to cost and availability of inference resources. However, we note that similar techniques for creating audio-caption pairs can be applied to other LLMs on the market.

\subsubsection{Data Postprocessing}
We apply additional postprocessing steps to refine the quality of generated captions. We filter out captions by setting a minimum/maximum threshold for word count. This ensures that short captions that lack information are not included, while excluding excessively wordy captions that tend to include unnecessary details or reflect poor grammar or sentence structure. Furthermore, we manually inspected a randomly selected subset of captions to ensure quality. 

\subsubsection{Dataset Analysis}
\begin{table*}[]
\centering
\caption{Comparison between AudioSetMix and existing audio-language datasets}
\begin{tabular}{c | c | c | c | c }
Dataset & Num. Audio-Caption Pairs & Duration (h) & Avg. Caption Length  & Perplexity\\
 \hline
\hline
AudioCaps  & 52904 & 144.94 & 4.01 & 1007.25\\
Clotho & 29645 & 37 & 16.97 & 296.76\\
MACS & 17685 & 9.83 & 7.01 & 1174.34\\
\textbf{AudioSetMix}~(Ours) & 49971 & 138.81 & 20.04 & 286.46
\end{tabular}
\label{tab:datastats}
\end{table*}
Table~\ref{tab:datastats} shows a comparison of statistics between AudioSetMix and human-annotated audio-language datasets. We note that because up to 5 distinct audio events can occur in a single AudioSetMix clip, a more complex caption is needed to fully describe the entire clip. Thus, the average caption for AudioSetMix is longer than the other datasets. To evaluate the quality of our captions compared to human-generated captions, we use GPT2~\cite{gpt2} to compute the average perplexity of captions. We observe that our captions have lower perplexity comparing to Clotho.

\section{Experiments}

In this section, we study the effectiveness of AudioSetMix for learning text-to-audio retrieval task. We additionally study the impact of AudioSetMix for improving model understanding of audio event modifiers, words that describe an attribute of an audio event such as volume or pitch. We provide a description of each task, as well as the experimental settings, results, and analysis. For all experiments performed, we use 16kHz sampling rate and 64-dimensional logmel-spectrogram with 1024-point Hamming window and 160 hop size to compute the audio input. 

\subsection{Text-to-Audio Retrieval}

\begin{table*}[]
\centering
\caption{R@K on Clotho test set}
\resizebox{\textwidth}{!}{
\begin{tabular}{l | l | c | c | c }
Model & Dataset & R@1 & R@5 & R@10 \\
\hline
\hline
ResNet38-BERT-medium & Baseline & 11.349 & 28.784 & 39.387  \\
ResNet38-BERT-medium & Baseline + AudioSetMix & \textbf{13.454} & \textbf{34.717} &  \underline{47.559}  \\
ResNet38-BERT-medium & Baseline+AudioSetMix+Hard Negatives& \underline{13.416} & \underline{34.392} & \textbf{48.306}  \\
\hline \hline

ResNet38-BERT-base & Baseline &  10.755 & 27.253 & 38.794  \\
ResNet38-BERT-base & Baseline+AudioSetMix & \underline{13.110} & \underline{34.928} & \textbf{48.631}  \\
ResNet38-BERT-base & Baseline+AudioSetMix+Hard Negatives & \textbf{13.454} & \textbf{35.138} & \underline{48.325}  \\
\hline\hline

ReSNet38-RoBERTa-large & Baseline & 11.770 & 29.971 & 41.665 \\
ReSNet38-RoBERTa-large & Baseline+AudioSetMix & \textbf{12.669} & \textbf{34.449} & \textbf{47.157} \\
ReSNet38-RoBERTa-large & Baseline+AudioSetMix+Hard Negatives & \underline{12.478} & \underline{33.071} & \underline{46.870} \\

\end{tabular}
}
\label{tab:retrieval}
\end{table*}
Text-to-audio retrieval involves retrieving the audio clip that best matches a given text caption/query from a database of clips. Retrieval is generally done by pushing matching audio-caption pairs closer in an embedding space and keeping non-matching pairs apart~\cite{wavcaps}.

\begin{figure}
    \centering
    \includegraphics[width=0.7\linewidth]{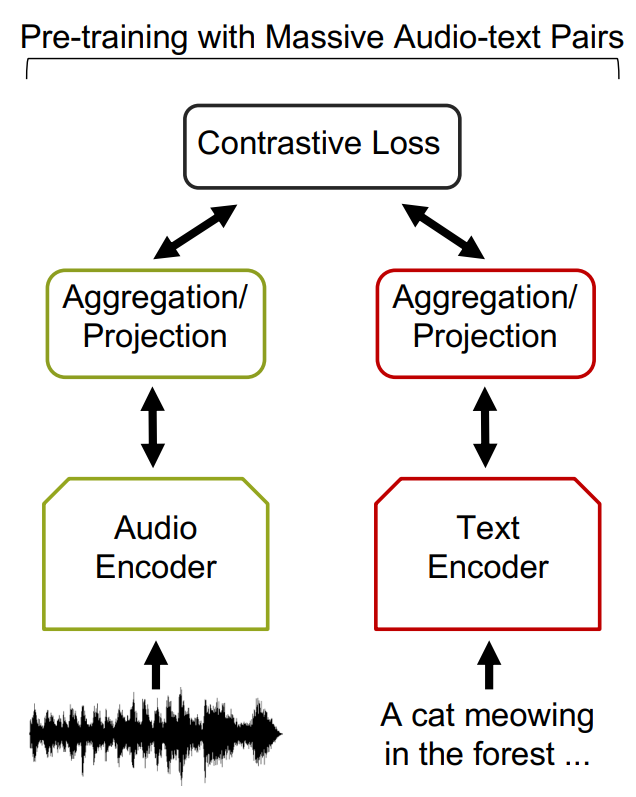}
    \caption{The baseline bi-encoder model architecture (from~\cite{leverage}).}
    \label{fig:base1}
\end{figure}

\subsubsection{Baseline Models}
Following ~\cite{leverage}, we select a common dual encoder architecture used in the 2022/2023 DCASE Challange on Language-Based Audio Retrieval~\cite{base} (Figure~\ref{fig:base1}). This architecture consists of audio encoder $E_a$ and text encoder $E_t$. For an audio-caption pair (A, T), the we computes an audio and text embedding, and project them to a shared dimension using a linear layer:
    
\begin{equation}
\begin{split}
    a &= \text{Proj}_a(E_a(A)), \\
    t &= \text{Proj}_t(E_t(T)).
\end{split}
\end{equation}

Next, assume we have a training batch of $B$ audio-caption pairs $(A_1, T_1), (A_2, T_2), ... (A_B, T_B)$. We denote the model's output for the $i^{th}$ pair $(A_i, T_i)$ as $a_i$ and $t_i$, respectively. We now compute a similarity score $s_{ij}$ between the ith audio clip and the jth caption using dot product:
\begin{equation}
\label{eq: sim}
\begin{split}
    s_{ij} = a_i \cdot t_j^T.
\end{split}
\end{equation}

We use the popular InfoNCE ~\cite{infonce} contrastive loss function to train the model. However, because an audio clip can potentially to multiple text captions in the training dataset, we wish to avoid penalizing the model for correctly associating these audio-text pairs when they occur in the same minibatch. Thus, following ~\cite{ilharcoloss}, we introduce a $B \times B$ masking term M where $B$ is the batch size:

\begin{equation}
\label{mask}
M_{ij} = \begin{cases}
        0, \text{ if $i^{th}$ clip matches $j^{th}$ caption} \\
        1, \text{ otherwise}. \\
        \end{cases}
\end{equation}

We slightly modify the InfoNCE loss function with learnable temperature $\tau$ using $M$, as shown in Eq~\ref{eq:loss}. We note that when $M$ consists of all ones, Eq~\ref{eq:loss} is identical to the standard InfoNCE loss.

\begin{equation}
L_{TA} = -\frac{1}{B}\sum_{i = 1}^{B} \log \frac{e^{\mathbf{s}_{ii}/\tau}}{\sum_{j = 1}^{B} \mathbf{M}_{ij} e^{\mathbf{s}_{ij}/\tau}} ,
\end{equation}
\begin{equation}
L_{AT} = -\frac{1}{B}\sum_{j = 1}^{B} \log \frac{e^{\mathbf{s}_{jj}/\tau}}{\sum_{i = 1}^{B} \mathbf{M}_{ij} e^{\mathbf{s}_{ij}}/\tau},
\end{equation}
\begin{equation}
\label{eq:loss}
L = L_{TA} + L_{AT}.
\end{equation}

Following ~\cite{base}, we select a \textit{ResNet38} model with pretrained weights from PANNs~\cite{panns} as $E_a$. To investigate whether more powerful text encoders are better at capturing the presence of audio event modifiers in the captions, we select BERT-medium~\cite{bert_medium}, BERT-base~\cite{bert}, and RoBERTa-large ~\cite{roberta} as our choices for $E_t$. 

\subsubsection{Text-to-Audio Retrieval Training}
Following ~\cite{base}, we combine training sets from multiple sources to form a single, larger training set. We selected AudioCaps ~\cite{audiocaps}, Clotho ~\cite{clotho}, and MACS~\cite{macs}. Because AudioCaps data consists of YouTube videos may become unavailable over time, we obtain 49k audio-caption pairs out of the original 50k. Combining these datasets gives us a total of 89k training audio-text pairs. We refer to this dataset as the baseline dataset

We trained our models with the loss function defined in Eq~\ref{eq:loss} for $20$ epochs using $10^{-5}$ learning rate, batch size of $64$, Adam optimizer, and cosine decay learning rate scheduler.   

We evaluate our models using the Recall@K metric (R@K), which is defined as follows: Given a dataset of audio-caption pairs, we first compute embeddings for caption and audio clip using the pretrained models respectively. Next, for each caption, we compute the similarity between its embedding and all audio clips using Eq~\ref{eq: sim}. The Recall@K metric is defined as the probability that the top $k$ most similar audio clips contains the targeted ground-truth clip. We use the test split from Clotho as the evaluation dataset. We report the recall for different $k$ for our baseline models in Table~\ref{tab:retrieval}. 

\subsubsection{Evaluating Modifier Understanding for Text-to-Audio Retrieval}
\begin{table*}[]
\centering
\caption{R@10 for original and flipped captions in MTe, separated by modifier category}

\resizebox{\columnwidth}{!}{
\begin{tabular}{c| c | c | c | c }

Caption Type & Duration & Pitch & Speed & Volume \\
\hline
\hline
Flipped  & 96.43 & 90.91 & 74.08 & 67.33 \\
Original & 98.21 & 88.63 & 74.81& 69.34 \\
\end{tabular}}
\label{tab:exp0}
\end{table*}
As shown in~\cite{leverage}, existing audio-language models depend heavily on the keywords in the caption, which are typically nouns and verbs.~\cite{leverage} finds that this over-reliance on keywords causes current audio-language models to not capture the order of the audio event. Motivated by~\cite{leverage}, we extended, and investigated whether audio-language models also fail to ``understand'' the modifiers for events, such as $loud$ vs. $quiet$. We study the impact from four categories of common modifiers \{volume, pitch, speed, and duration\} for audio events, to the models. This is aligned with the methods to produce the AudioSetMix in Section~\ref{sec:datagen}. To do this, we create a subset from the Clotho and AudioCaps evaluation sets called ``Modifier Test Set'' (MTe). Captions in MTe contain words describing one of these modifiers, such as $loud$, $slow$, $short$ and etc.. In total, this gives us $700$ pairs of data in MTe. 

We first determine if existing audio-language models can capture modifiers in the caption. For this experiment, we select the ResNet38-Bert-base model as our baseline. Following the BAT test introduced in~\cite{leverage}, we replace each modifier in MTe with an antonym, forming what we call the flipped caption. For instance, the modifier $loud$ would be replaced with $quiet$, and the modifier $quickly$ would be replaced with $slowly$. We then use the flipped captions to retrieve the original audio and report the recall as performance metrics. The retrieval set is the set of all other audio-text pairs containing the same class of modifier. If the model is able to distinguish modifiers, we would expect that the recall scores would degrade significantly with ``flipped caption'' in comparison to using the original captions as the query. The results are reported in Table~\ref{tab:exp0}. We see only a marginal change in performance, which suggest that the models failed to learn to distinguish the modifiers well.

To more rigorously study model understanding of modifiers, we design the Modifier Understanding Test (MUT). For each audio-caption pair in MTe, We first compute the embedding distances between the original audio and both the flipped and original captions. We then count the percentage of times that the flipped caption embedding is closer to the audio embedding than the original caption embedding. If the model completely fails to capture the presence of modifiers, we should expect that the flipped caption is closer 50\% of the time. On the other hand, a perfect model should have a score of 0\%. As shown by the results in Table~\ref{tab:exp1}, the model unsurprisingly performs significantly better than random choice in every modifier category, but is far from perfect.

\begin{table*}[]
\small
\centering
\caption{Performance on MUT for model trained on the baseline dataset, the augmented dataset, and the augmented dataset with hard negatives. Performance is measured as the percentage of samples for which the flipped text caption is closer to the audio than the original caption (lower is better).   }
\begin{tabular}{l | l | c | c | c | c }
Model & Dataset & Duration & Pitch & Speed & Volume \\
 \hline
\hline
ResNet38-BERT-medium & Baseline  & 42.857 & \textbf{40.909} & 46.667 & 44.889 \\
ResNet38-BERT-medium & Baseline+AudioSetMix & \underline{35.714} & \underline{43.181} & \underline{36.296} & \underline{37.875}\\
ResNet38-BERT-medium & Baseline+AudioSetMix+Hard Negatives & \textbf{33.928} & 45.450 & \textbf{32.592} & \textbf{35.871} \\
\hline
\hline
ResNet38-BERT-base & Baseline  & 44.643 & \underline{31.818} & 38.519 & \underline{33.066} \\
ResNet38-BERT-base & Baseline+AudioSetMix & \underline{37.50} & 38.636 & \underline{33.334} & 33.266 \\
ResNet38-BERT-base & Baseline+AudioSetMix+Hard Negatives & \textbf{35.714} & \underline{31.818} & \textbf{31.111} & \textbf{28.857} \\
\hline
\hline
ResNet38-RoBERTa-large & Baseline  & 48.214 & \textbf{40.909} & 39.259 & 36.272 \\
ResNet38-RoBERTa-large & Baseline+AudioSetMix & \underline{46.428} & \underline{47.727} & \textbf{35.560} & \textbf{31.863} \\
ResNet38-RoBERTa-large & Baseline+AudioSetMix+Hard Negatives & \textbf{37.50}  & 50.0 & \underline{38.518} & \underline{32.865} \\
\end{tabular}
\label{tab:exp1}
\end{table*}

Finally, we wish to evaluate the model's ability to distinguish between different categories of modifiers. We perform retrieval on MTe using the original captions as queries where the retrieval set is the set of every audio-text pairs in MTe, and report the recall scores in Table ~\ref{tab:exp2}. We refer to this experiment as Modifier Differentiating Test (MDT).

\begin{table*}[]
\centering
\caption{Performance on MDT for model trained on the baseline dataset, the augmented dataset, and the augmented dataset with hard negatives.}
\begin{tabular}{l| l | c | c | c  }
 Model & Dataset & R@1 & R@5 & R@10\\
 \hline
\hline
ResNet38-BERT-medium & Baseline  & \underline{15.258}  & 47.002  &  59.809 \\
ResNet38-BERT-medium & Baseline+AudioSetMix  & 15.208 & \underline{53.678} & \underline{67.438} \\
ResNet38-BERT-medium & Baseline+AudioSetMix+Hard Negatives & \textbf{16.485}  &  \textbf{56.675} & \textbf{68.801}\\
\hline
\hline
ResNet38-BERT-base & Baseline      & 15.122  &  47.002 &  61.580 \\
ResNet38-BERT-base & Baseline+AudioSetMix  & \underline{16.893}  & \underline{55.722}  & \underline{70.163} \\
ResNet38-BERT-base & Baseline+AudioSetMix+Hard Negatives & \textbf{17.029} & \textbf{55.858} & \textbf{70.980}\\
\hline
\hline
ResNet38-RoBERTa-large & Baseline  & 14.032  &  47.138 &  61.989\\
ResNet38-RoBERTa-large & Baseline+AudioSetMix  &  \textbf{16.076}  & \textbf{55.585} &  \textbf{69.891} \\
ResNet38-RoBERTa-large & Baseline+AudioSetMix+Hard Negatives & \underline{15.122}  & \underline{53.814} & \underline{68.528} \\
\end{tabular}
\label{tab:exp2}
\end{table*}

\subsubsection{Training with AudioSetMix}
The percentage of sentences in the training data that contain modifiers is extremely small, as shown in Table ~\ref{tab:stats}. As such, we study whether increase the number of modifiers in the training data improves understanding of modifiers. We augment the baseline dataset using AudioSetMix, giving us a total of 132k training audio-text pairs. We train new models using the same training procedure as the baseline.  As shown in Table~\ref{tab:exp1}, the models show significant gains in understanding duration, speed, and volume modifiers when trained with AudioSetMix. Furthermore, Table~\ref{tab:exp2} shows that all models improve in their ability to distinguish between the different modifier categories.

\begin{table}[]
\centering
\caption{Top: Number of sentences in MTe containing each modifier type, with \% indicating size relative to the full test set. Bottom: the same distribution for pretraining set, with \% indicating size relative to the full pretraining set.}
\resizebox{\columnwidth}{!}{
\begin{tabular}{c | c| c | c | c  }
 Dataset & Duration & Pitch & Speed & Volume\\
 \hline
\hline
MTe          &  56 (0.5\%) & 44 (0.4\%)  &  135 (1.3\%) & 499 (4.8\%)\\
Training     &  452 (0.5\%)  & 347 (0.3\%)  & 778 (0.8\%) & 4540 (5.1\%)\\
\end{tabular}}
\label{tab:stats}
\end{table}

\subsubsection{Training with Generated Hard Negatives}
In contrastive learning, each audio clip $A_i$ is contrasted with other texts $T_j$, which usually describe completely different clips. As such, models are able to ignore finer details of modifiers in clips and simply focus on audio events~\cite{videotime}. We hypothesize that hard negative examples that contain the same audio events are needed to encourage models to capture and understand modifiers. In contrast to~\cite{wavcaps}, our data generation pipeline Figure~\ref{fig:pipeline} provides a natural way to generate hard negatives for each data point in AudioSetMix. Recall that in the data generation process, we sample a set of audio clips $A$ for augmentation. We randomly select a set of augmentation operations $T$ for $A_i \in A$, with each $T_j\in T$ applies to $A_i$, and produced the augmented clips $C_i$. We finally assemble the audio clips in $[C_i]$ by mixture of concatenation and mixing. For hard negatives, while keeping the same procedure as outlined above, we ``reverse'' each operation $T_j$ (i.e. if the original $T_j$ is $loud$, the new transformation $T^{\prime}_j$ is $quiet$). We argue that this creates hard negative samples since the two final augmented clips contain the same set of audio events, but with opposite augmentation operations applied to each. This makes it more challenging for the model to learn to distinguish them and forces the model to attend to the audio modifiers. 

To train with these hard negatives, we randomly select $c$ AudioSetMix inputs in each minibatch and generate their hard negatives. We then append the hard negatives to the minibatch and perform the model update. We empirically find that $c=16$ works well for our batch size setting of 64. In Table ~\ref{tab:exp1}, we show that model performance on MUT is generally improved using hard negatives on all modifier categories except pitch. We hypothesize that pitch may be difficult to improve upon because pitch is not as broadly applicable of a modifier, meaning that pitch augmentation in AudioSetMix may not be meaningful semantically (i.e. applying a high-pitch transformation to a base sound of ``tree falling over''). 

Finally, we compare our models trained on AudioSetMix and hard negatives with the baseline models on the original text-to-audio retrieval task. We evaluate each model using the evaluation set from Clotho and show the results in Table ~\ref{tab:retrieval}. We note that the models trained using AudioSetMix and hard negatives beat the baseline results consistently on Clotho. In contrast to the previous experiments, the addition of hard negative mining provides only a marginal improvement to the recall scores. Because a) the hard negatives are only concerned with modifiers and b) the lack of modifiers in the eval sets, we hypothesize that the benefits of hard negative mining are not observable in this experiment. 

\section{Conclusion}
In this work, we introduce AudioSetMix, a weakly-labelled audio-caption pair dataset created by applying audio transformations to existing datasets. We propose a pipeline to augment/combine audio clips and generate a corresponding caption using LLM. We evaluate AudioSetMix on text-to-audio retrieval and demonstrate that AudioSetMix improves model understanding of audio event modifiers. In future we hope to evaluate models trained from AudioSetMix using human feedback.

\textbf{Acknowledgements} We thank David Harwath for providing insight and expertise that greatly assisted the research in this work. 

\newpage
{\small
\bibliographystyle{plain}
\clearpage
\bibliography{refs}
}
\end{document}